\documentstyle[12pt,aasms4,graphicx]{article}
\begin{document}

\title{
Is the Fast Evolution Scenario for Virialized Compact Groups Really
Compelling? The Role of a Dark Massive Group Halo}

\lefthead{Role of Dark Halos in the evolution of CGs}

\author{M.A. G\'omez-Flechoso\altaffilmark{1} and R. Dom\'{\i}nguez-Tenreiro} 
\affil{Dpt. F\'{\i}sica Te\'orica C-XI, 
Universidad Aut\'onoma de Madrid, \\
E-28049 Cantoblanco, 
Madrid, 
Spain}

\altaffiltext{1}{Present address: Observatoire de Gen\`eve,
Ch. des Maillettes 51, Ch-1290 Sauverny (Switzerland)}

\begin{abstract}

We report on results of N-body simulations aimed at testing
the hypothesis that galaxies in X-ray emitting (i.e., virialized)
Compact Groups are not tidally stripped when they are embedded in a
common, massive, quiescent dark matter halo.
To disentangle the effects of interactions from
spurious effects due to an incorrect choice of the initial
galaxy model configurations, these have been chosen to be
tidally-limited King spheres, representing systems in
quasi-equilibrium within the tidal field of the halo.
The potential of the halo has been assumed to be frozen
and the braking due to dynamical friction neglected. 
Our results confirm the hypothesis of low rates of tidal stripping 
and suggest a scenario for virialized Compact Group evolution in their
quiescent phases with only very moderate tidally induced galaxy evolution
can be generally expected. 
This implies the group stability, provided that
the dynamical friction timescales 
in these systems are not much shorter
than the Hubble time. We discuss briefly this possibility, in
particular taking account 
of the similarity between the velocity
dispersions of a typical virialized Compact Groups and the internal
velocity dispersion of typical
member galaxies.
A number of puzzling observational data on Compact Groups
can be easily explained in this framework.
Other observations would be better understood as the
result of enhanced merging activity in the proto-group
environment, leading to virialized  Compact Group
formation
through mergers of lower mass halos,  as predicted by
 hierarchical scenarios of structure formation.

\end{abstract}


\section{INTRODUCTION}
\label{intro}
Compact groups (CG) of galaxies are relatively isolated systems composed
of a small number of galaxies (three or more) forming configurations
with very low median projected intergalactic separation,
providing the highest projected galaxy number density systems in the sky.
 Hickson (1982) has pioneered
the use of specific, quantitative selection criteria to create CG catalogs.    
Other catalogs now available include the Southern CG catalog 
(Prandoni, Iovino \& MacGillivray 1994), and those produced from the CfA2 (Barton et al. 1996)
and
Las Campanas (Allam \& Tucker 1998) redshift surveys.

 Groups in Hickson's catalog and their member galaxies have been observed
at many wavelengths and the extensive available data pose
a number of challenging questions.
Their high galaxy number density and relatively low
 velocity dispersion would imply,
on theoretical grounds, short dynamical timescales and rapid evolution,
making CGs the ideal sites for galaxy tidal interactions (and, eventually,
tidal disruption) and mergers to occur. These would result in the
appearance of more frequent dynamical peculiarities in CG galaxies,
as well as enhanced star formation rates and nuclear activity compared
with galaxies in less dense environments.
Enhanced  galaxy merging activity  would produce merger remnants and would lead
to the formation of blue luminous ellipticals in a time scale of 
$\sim$ 1 Gyr. Tidal disruption would cause, eventually,  the CG  
 disappearance in a few crossing times ( $\sim
1.5 $ Gyr; Barnes 1989). However,
observations of Hickson Compact Groups (HCGs) do not
support these predictions.
A fraction as high as 43$\%$ of galaxies in HCGs may be interacting
(Mendes de Oliveira \& Hickson 1994), but no correlation has been found between
the number of interacting galaxies in a given group and the group global
parameters (velocity dispersion, crossing times, X-ray properties;
 Mendes de Oliveira \& Hickson 1994; Pildis, Bregman \& Schombert 1995).
Also, a high fraction of spirals in HCGs have asymmetric and peculiar
rotation curves, but these dynamical peculiarities do not
correlate with the CG global properties, even if the fraction
of galaxies with peculiar rotation curves is higher 
among those belonging to CGs with lower galaxy velocity dispersion inside the group,
$\sigma_{\rm group} \leq 100$ km s$^{-1}$
  (Nishiura et al. 2000).
The $H\alpha$, CO and FIR observations of the
HCG galaxies have shown neither enhancement in their
present-day star formation rates nor in their
star formation histories, when compared with control samples of
field galaxies (see Verdes-Montenegro
et al. 1998;  Iglesias-P\'aramo \& V\'{\i}lchez 1999 and references
therein). 
Coziol et al. (1997), Coziol, Iovino \& de Carvalho (2000) and Allam et al. (1999)
have found depressed star formation,
relative to control samples, in CGs of the Hickson, Southern and Las Campanas
catalogs, respectively, and,
 also, a density-morphology-activity
relation in both HCGs and Southern CGs.
Moreover, even if the fraction of galaxies in HCGs
that are merger remnants is still controversial, observations suggest
that it is not high  
(only $\sim 7\%$  in Zepf 1993; 
see also Hickson 1997 and references therein).
In HCGs, the first ranked galaxy sample and the global population have the same 
fraction of E or S0 galaxies (Hickson 1982).
Observations also suggest that there are not
more blue ellipticals in HCGs than in other environments (Zepf \& Whitmore 1991). 
  HCGs have a relatively
large spiral fraction, but it is anticorrelated with the galaxy 
velocity dispersion (Hickson, Kindl \& Huchra 1988) 
and is lower in 
X-ray luminous CGs  (these two parameters are, on their turn, correlated,
with X-ray detected CGs having $\sigma_{\rm group} \geq 100$ km s$^{-1}$,
see  Mulchaey 2000).
 These results and 
the HCG number density are in contradiction with the theoretical framework
of rapid evolution, supported by the first N-body simulations on CG
evolution (Barnes 1985, 1989; Mamon 1987).
To surmount these difficulties, some authors have pointed out that a large 
fraction of HCGs are not real dense configurations, but chance superpositions
 of galaxies  (Mamon 1986;
Diaferio, Geller \& Ramella  1994; Hernquist,
Katz \& Weinberg 1995, but also see Hickson \& Rood 1988). 
But this scenario does not account for the existence
of a diffuse X-ray emission from hot intergroup gas in 75$\%$ HCGs
(Ponman et al. 1996).
This gas has been found to have almost the same temperature in 
 HCGs 
 where it has been detected ($T_{\rm X} \simeq $0.9 keV,
with a detection limit of $T_{\rm X} \simeq $0.3 keV).
The projected
gas density profiles are consistent with
$\Sigma_{\rm gas}(s) = \Sigma_{\rm gas}^0 / [1 + (s/R_c)^2]^{(3\beta_{\rm
gas}-1)/2}$,
with core radius
$4 \leq R_c \leq 30 \, h^{-1} \, \rm kpc$ and slope
$0.38 \leq \beta_{\rm gas} \leq
0.92$ for HCGs (see Mulchaey et al. 1996 and references therein).
Assuming hydrostatic equilibrium
($\beta$-model, see Cavaliere 1973), these observations imply a total
 group binding mass of $M_{\rm group} \simeq 2 \times 10^{13}$ M$_{\odot}$ at
$R \sim 300$ kpc for groups of the HCG catalog 
(see Mulchaey et al. 1996, and references therein),
 implying that  galaxies in these CGs are embedded
in a common massive, concentrated  dark matter halo.

We present in this Letter a new framework to understand 
X-ray emitting CGs of galaxies as
stable systems against tidal disruption of their
galaxies, in the quiescent phases of their evolution\footnote[1]{In 
any hierarchical scenario for structure
formation, violent (i.e., merging) phases and quiescent phases follow
each other in the evolution of any astronomical system.
CGs halos would have been formed through merging 
of lower mass halos.}.
 The main hypothesis is that 
the common virialized massive concentrated 
dark matter halo, in which galaxies are embedded
in CGs, stabilizes 
both the group and the individual galaxies,
so they are able to remain in
dynamical equilibrium during many group crossing times
after halo formation. This possibility has 
been tested through a series of N-body simulations, whose results confirm
this assumption. 

\section{A SCENARIO FOR STABLE CGS OF GALAXIES}
\subsection{Description of the simulations}
To study the stability of the galaxies in CGs
and the group itself, we have built 
initial model galaxies and left them to orbit in the
external halo potential.
We have considered that the massive dark matter
halos in which galaxies are
embedded are consistent with the gas distribution ($\beta$-model) 
as explained above.
A statistical analysis of  the  distribution of projected
galaxy pair separations in accordant
redshift quartets of HCGs by Montoya et al. (1996)
shows that galaxy density profiles are consistent with those of the X-ray
emitting gas (see above), with slope parameter $ \beta_{\rm gal}$
instead of $ \beta_{\rm gas}$. In this way, 
the galaxy and halo mass  distributions are
determined by four parameters among the
following: $M_{\rm group}$, $R_c$, $ \beta_{\rm gas}$, $T_{\rm X}$ (given
by X-ray data), slope of the galaxy distribution profile, $ \beta_{\rm gal}$,
and galaxy velocity dispersion inside the group, $\sigma_{\rm group}$. We recall
that, solving the Jeans equation, the $\beta$-model 
predicts that $\beta_{\rm spec} = 
\sigma_{\rm group}^2/(kT_{\rm X}/\mu m_p)
= \beta_{\rm gas} / \beta_{\rm gal}$, and indeed this is the case for HCGs
(Montoya et al. 1996).
To illustrate the stabilizing role of massive halos,
 results  for two different halo masses are reported:
CG1 group model, with $M_{\rm group} = 1.85 \times 10^{13}$ M$_{\odot}$, and 
CG2 group model, with $M_{\rm group} = 2.78 \times 10^{12}$ M$_{\odot}$ at 
$R=300$ kpc. The core radius of the group is $R_c=24$ kpc (Mulchaey et al. 1996).
The remaining parameters of the halos (Table \ref{tbl1}) 
are also consistent with those of HCGs, taking $H_0=50$ km s$^{-1}$ Mpc$^{-1}$ 
(Hickson et al. 1992).

The initial positions and velocities of the galaxies in the halo, 
$R_{i, in}$ and $V_{i, in}$, $i = 1,2,...,$N$_{\rm G}$
(N$_{\rm G}$ is the number of galaxies in the group;
as very often accordant HCGs are quartets, we have taken N$_G$=4)  
have been assigned in the following way: (a) the positions are
Monte-Carlo realizations of the average HCG galaxy number density profile
found by Montoya et al.
(1996), and (b) the velocities are assigned randomly with the condition that
$\sum_{i=1}^{\rm N_G} V_i^2 ={\rm N_G}\sigma_{group}^2$. 
The $R_{i, in}$, $V_{i, in}$ and
the orbital angular momentum, $L_i$, are 
given in Table \ref{tbl2}.
Two sets of galaxy initial positions and velocities are
reported for the massive CG1 group model; for one of them
(CG1M), they have been prepared
in such a way that they lead to a binary merger event of two galaxies.

Galaxies have been assumed to be self-gravitating
configurations in quasi-equilibrium with the environment,
that is, with the tidal
field of the halo \footnote[2]{Note that to properly quantify environmental effects,
it is necessary that the initial model represents a galaxy
in quasi-equilibrium within the tidal field.
Otherwise, it would be difficult to disentangle between the effects that are
effectively due to interactions and those that are spurious,
due to an incorrect choice of the initial galaxy configuration,
see G\'omez-Flechoso and Dom\'{\i}nguez-Tenreiro (2001b, in preparation) for a discussion. In fact, galaxy infall in numerical 
simulations of small galaxy group dynamics is mostly caused
by (spurious) tidal heating, enhanced in some cases by two-body
numerical heating.}.
As the simplest choice for the initial galaxy model, we 
assume that they are spheres with an isotropic velocity dispersion tensor.  
In this case, the tidal field determines the limiting or tidal radius,
$r_{\rm t}$, of the configuration (King 1962; 
G\'omez-Flechoso \& Dom\'{\i}nguez-Tenreiro 2001, hereafter GD01),
leading to the so-called tidally-limited ($t$-limited) King spheres,
 based on the King-Michie velocity distribution function
 (King 1966; Michie 1963).
These $ t$-limited King spheres are characterized
by two free parameters among the total galaxy mass, $M_{\rm gal}$,
the 1-dimensional internal velocity dispersion, $\sigma_{\rm gal}$, the galaxy
core radius, $r_{0}$, and the central potential, $W_0$
(see GD01 for details).
Note that the widely used standard King spheres,
whose limiting radius is left free (and not, as due, fixed
by the tidal field), have instead three free parameters.
In GD01 
we describe a method to build $ t$-limited King spheres with
prefixed values of $M_{\rm gal}$ and $r_{0}$.
The initial configurations for our CG galaxies are Monte-Carlo
realizations of the $ t$-limited King spheres,
with N$_{\rm p}$=10000 particles,
$M_{\rm gal} = 1.1 \times 10^{11}$ M$_{\odot}$ and $r_{0}=0.2$ kpc.
In Table \ref{tbl2} we give the tidal radius, $r_t$, and the 
$\sigma_{\rm gal}$ and $W_0$ parameters corresponding to each galaxy model
of the  CGs reported in this Letter.
 The values of these parameters are within their observational range. 
These CG models have been left to evolve during a time interval 
of  10$^{10}$ years ($\simeq 10^2$
group crossing times). Evolution has been followed by integrating
the equation of motion for each constituent particle in the 
combined potential of the other particles and the halo
 by means of a vectorized version of the treecode (Hernquist 1987).
 As we are interested in galaxy stability against tidal
disruption, the halo potential has been taken to be smooth,
so dynamical friction effects have not been considered.
However, no important dynamical friction effects
are expected (see $\S$\ref{summary}).

\subsection{Results}
For the massive type CG1 group,
both the individual galaxies and the group itself are 
stable against tidal stripping,
therefore very low tidally induced evolution can be expected
in any of them. By galaxy stability we mean that:
(a) the individual galaxies do not lose appreciably mass, as
the fraction of mass stripped by tidal forces is low ($5-25\%$,
depending on the galaxy initial position),
(b) the mass density profiles of the individual galaxies 
do not change, and,
(c) the velocity distribution function of each galaxy does not evolve 
significantly, and, therefore, their internal velocity 
dispersion, $\sigma_{\rm gal, i}$, is almost constant; 
only a mild velocity anisotropy is developed in each galaxy.
The CG does not change either, that is:
(a) galaxies do not end up at the center of the configuration; on the contrary,
the  values of the  galaxy pericentric and apocentric distances,
 as well as the median intergalactic
separation, remain stable (see Figure 1a),
 and,
(b) the galaxy velocity dispersion of the group, 
$\sigma_{\rm group}$, does not change on average (Figure 1b).
The same qualitative results have been found for the CG1M group,
where the merger of the A and C galaxies soon after the
begining of the simulation does not affect
either the dynamical properties of the other galaxies
in the group, nor the overall group evolution (see Figure 1).

All these results are mostly determined by the choice of the initial
galaxy model configurations with an average density 
consistent with the halo density at galaxy pericenter,
that is, they are tidal quasi-equilibrium 
solutions in the tidal field,
and they are roughly independent on the galaxy or halo
density profiles (see GD01 for details).

As  $M_{\rm group}$ decreases,
binary interactions become more important
relative to the overall halo potential. As a consequence, galaxies
orbiting inside this halo become more likely to be
tidally stripped and lose mass,
and they tend to fall towards the halo center,
where their remnants eventually end up, merging with each other
and causing the group tidal disruption.
An example of such a process is provided by
the evolution of the low halo mass CG2 group model, where  
half of the mass inside 100 kpc is in the galaxies.
The galaxies of the CG2 group lose $30-50 \%$ of their mass during the first
4 Gyrs of the simulation. Their remnants merge 
 and fall to the halo center.
Figure 1 illustrates the group disappearance as a consequence of these mass
loss. 
Taking more massive galaxies in this last experiment would 
result in a faster group disappearance, because the determining point here
is the importance of the galaxy-galaxy interactions relative to the
galaxy-halo interactions.

\section{SUMMARY AND DISCUSSION}
\label{summary}
According to current  scenarios of hierarchical structure
formation, dark matter halos form through violent merger 
events from previously formed lower mass halos.
These halos host baryonic clumps (either normal galaxies, 
dwarfs, and possibly HI clouds) that 
survive the halo merging and virialization, 
and, after this process is completed,
they orbit inside the newly formed halo (Navarro, Frenk \& White 1995, hereafter
NFW;
Tissera \& Dom\'{\i}nguez-Tenreiro 1998).
In the present Letter, we show that very little mass is 
stripped from typical galaxies orbiting inside the potential
well of a quiescent, massive CG-like 
halo,  provided that the halo mass is 
 similar to that inferred from observations of
CGs with diffuse X-ray emission and that the initial galaxy models
are self-gravitating configurations in quasi-equilibrium  
within the tidal field of the halo.
When these two conditions are met, the rate of tidal stripping is
roughly independent of the galaxy or halo density
profile, as shown by GD01, where a cuspy halo density
profile as those obtained by NFW
has been adopted. So, profiles do not need
to have low concentrations, as demanded by Athanassoula,
Makino \& Bosma (1997). These authors used standard (as opposed
to $t$-limited) King spheres as initial model galaxies,
that suffer from important mass loss
unless that the CG halos they orbit are almost homogeneous.
An important  consequence of the 
tidal quasi-equilibrium in astronomical systems is that
 dynamical timescales for
galaxy infall in CGs, and the ensuing CG disappearance,
 are not set by mass loss due to tidal stripping. 

Among other 
processes causing galaxy infall to the center of the
configuration the most efficient is dynamical friction,
whose timescales, $t_{\rm fric}$,
however, 
are difficult to predict.
The well known Chandrasekhar formula gives, for a $\sim 2 \times
10^{13}$ M$_{\odot}$ NFW halo and a $ \sim 10^{11}$ M$_{\odot}$ 
galaxy initially placed at $\sim$ 40 - 90 kpc from
the halo center, a value of $t_{\rm fric} \sim $ 3 - 6 Gyrs
(Klypin et al. 1999). This  situation corresponds roughly to the 
virialized CGs described in this paper.
Dom\'{\i}nguez-Tenreiro \& G\'omez-Flechoso (1998, hereafter DG98) have
shown that 
 if the internal velocity dispersion of the orbiting galaxies
is of the order of  $\sigma_{\rm group}$, then
$t_{\rm fric} $ could easily be a factor of $\sim$ 2 - 3
longer than the values above, and even much more longer, 
depending  very strongly on the particular values of the
GC halo and galaxy parameters.
Unfortunately, these analytical estimations of $t_{\rm fric} $
are difficult to test through numerical simulations
of the evolution of CGs with live halos.
The main pitfall is the correct implementation in the simulations of the 
ratio between the fluctuating forces
(causing dynamical friction)
and the smooth forces, $R_{\rm fs}$, that must be low
to accurately simulate the dynamical friction effects.
And so, the proper sampling of a virialized CG-like halo
would demand a too high number of particles 
for current computer facilities.
If $R_{\rm fs}$ is unphysically high, it results
into undesired numerical effects, leading to an overestimation
of the fluctuating force intensity and, consequently, to a too
low $t_{\rm fric} $ value. This is the so-called discreetness
effect (see Hernquist \& Weinberg 1989; see
also Eq. (14) in DG98).
The simulations of CG evolution by Bode, Cohn \& Lugger (1993)
suggest that $t_{\rm fric} $ is roughly consistent with
 Chandrasekhar's formula predictions; however, these simulations
could suffer from a discreetness problem, as they  use
a too low number of particles per galaxy and halo.  
In conclusion, we see that even if the case for
 inefficient dynamical friction
in CGs is not definitively closed,
 analytical estimations of $t_{\rm fric} $
suggest that,  
on theoretical grounds, 
the need of a rapid evolution scenario for virialized  CGs
becomes rather questionable.

A scenario where virialized CGs are dynamically quiescent
since the end of the violent relaxation phase ensuing
the merger episode that formed their massive dark matter halos
could explain a number of difficulties met
by the rapid evolution scenario (see $\S$\ref{intro}).

First, the lack of clear correlations between the rate of galaxy
interactions and CG global parameters
 is naturally explained if the galaxies are embedded
in a common massive halo that determines their trajectories.
The galaxy-galaxy binary interactions only represent 
perturbations to the overall halo potential and they are 
uncorrelated with global
CG parameters. As shown for the CG1M group, if 
one such binary interaction occurs, the remaining CG galaxies
are completely unaffected by this process.  
Binary interactions do destroy CGs if they are important relative to
the global field of forces caused by the halo, that is, if
the common halo is not massive
enough compared with the mass of individual galaxies (the CG2 group).
So, a range of CG halo masses
could be responsible for the different degrees of dynamical
evolutionary stages of CGs found
by Ribeiro et al. (1998). But these could also result from
different degrees of virialization after the merging 
event  involving 
galaxy halos and leading to the common massive dark CG halo. 

Second, the lack of enhanced merger activity {\it inside quiescent}
CG halos found in our simulations suggests that CG early-type galaxies
are unlikely to form in the quiet phases.
It explains
the anticorrelation found between the spiral fraction in CGs
and $\sigma_{\rm group}$ or the X-ray luminosity.
The CG early-type galaxies can be formed through the
same merger trees responsible for CG halo formation in a proto-group
environment. 
The hierarchical scenario predicts that the merger rates are
higher in dense environments than in isolation
 (Evrard, Silk \& Szalay 1990; Tissera \& Dom\'{\i}nguez-Tenreiro 1998;
 see also V\'azquez-Semadeni 1994), where systems form earlier,
and lead to more massive compact halos
(CGs would be more likely to appear within 
looser groups than isolated, see
Barton, de Carvalho \& Geller 1998) and to more early-type
galaxies within more massive (or more X-ray luminous or with
a higher $\sigma_{\rm group}$, given the correlations
among these quantities, see Mulchaey 2000) CG halos. 
These pre-virialization violent mergers could also have 
triggered important gas inflows in the baryonic systems,
giving rise, at that time, to starburst galaxies, that evolved
into low luminosity AGNs in the
post-merger phase, as the gas supply was gradually exhausted.
This could be the origin of the density-morphology-activity relation found
by Coziol et al. (1997, 2000), difficult to explain in
the framework of the replenishment scenario 
(Governato et al. 1996).

Finally, the lack of enhancement of star formation in the main
body of CGs as compared to field galaxies 
can be seen as another consequence of the
density-morphology-activity relation: starbursts would have
been induced sometime in the past, and the faster evolving
galaxy members, placed at the CG cores, are now observed
in a quiet phase, as are early-type galaxies in galaxy clusters.

It is a pleasure to thank l'Observatoire de Gen\`eve for its kind
hospitality while this work was finished.
MAGF was supported by the Direcci\'on General de
Ense\~nanza Superior (DGES, Spain) through a
fellowship. It also supported in part this work, grants AEN93-0673,
PB93-0252 and PB96-0029.

\clearpage

\begin{deluxetable}{lcccccc}
\tablecolumns{7}
\tablewidth{0pc}
\tablecaption{Group parameters}
\tablehead{
\colhead{} & \colhead{$\beta_{\rm gas}$\tablenotemark{a}\tablenotetext{a}{Typical values from X-ray observations (see Mulchaey et al 1996)}} & \colhead{$T_X$} & \colhead{$R_C$} & \colhead{$M_{\rm group}(<300 \,{\rm kpc})$} & \colhead{$\sigma_{\rm group}$} & \colhead{$\beta_{\rm gal}$\tablenotemark{b}\tablenotetext{b}{Best-fit value by Montoya et al. (1996)}}\\
\colhead{ }&  \colhead{ }         & \colhead{(keV)} &\colhead{(kpc)} & \colhead{(M$_{\odot}$)}   & \colhead{(km/s)} & \colhead{}}
\startdata
CG1 & 0.56 & 1.0 & 24.0 & $1.85\times 10^{13}$ & 252 & 1.4\\
CG2 &  0.42 & 0.2 & 24.0 & $2.78\times 10^{12}$ & 98 & 1.4\\
\enddata
\label{tbl1}
\end{deluxetable}
 
\clearpage

\begin{deluxetable}{lcccccccccccccc}
\tablecolumns{15}
\tablewidth{0pc}
\tablecaption{Galaxy parameters}
\tablehead{
\colhead{} &\multicolumn{4}{c}{CG1} &\colhead{\phantom{.}}& \multicolumn{4}{c}{CG1M} &\colhead{\phantom{.}}& \multicolumn{4}{c}{CG2}\\
\colhead{} & \colhead{A} & \colhead{B} & \colhead{C} & \colhead{D} &\colhead{}& \colhead{A} & \colhead{B} & \colhead{C} & \colhead{D} &\colhead{}& \colhead{A} & \colhead{B} & \colhead{C} & \colhead{D} }
\startdata
$\sigma_{\rm gal}$ (km/s) & 344 & 326 & 277 & 284 && 351 & 336 & 297 & 279 && 266 & 237 & 227 & 233 \\
$r_t$ (kpc) & 8.4 & 9.6 & 14.5 & 14.2 && 7.6 & 8.9 & 12.0 & 14.6 && 16.6 & 20.7 & 22.4 & 22.1\\
$W_o$ & 7.3 &7.5& 8.2& 8.1&&7.2& 7.4& 7.9& 8.1&& 8.3& 8.7& 8.9&8.8\\
$R_{i, in}$ (kpc) &32.7 & 42.6 & 72.6 & 88.1&&6.8&38.8&48.6&85.7&&29.7&38.7&45.8&52.5\\
$V_{i, in}$ (km/s)&497&465&234&495&&327&489&186&618&&181&79&253&107\\
$L_i$ (Mpc km/s)&15.5&18.8&6.0&37.4&&1.2&19.0&7.4&27.8&&4.4&2.9&2.1&5.4\\
\enddata
\tablecomments{The mass and the core radius of all the galaxies are $M_{\rm gal}=
1.1 \times 10^{11}$ M$_{\odot}$ and $r_0=0.2$ kpc, respectively.}
\label{tbl2}
\end{deluxetable}

\clearpage

\begin{center}
\includegraphics[width=14cm]{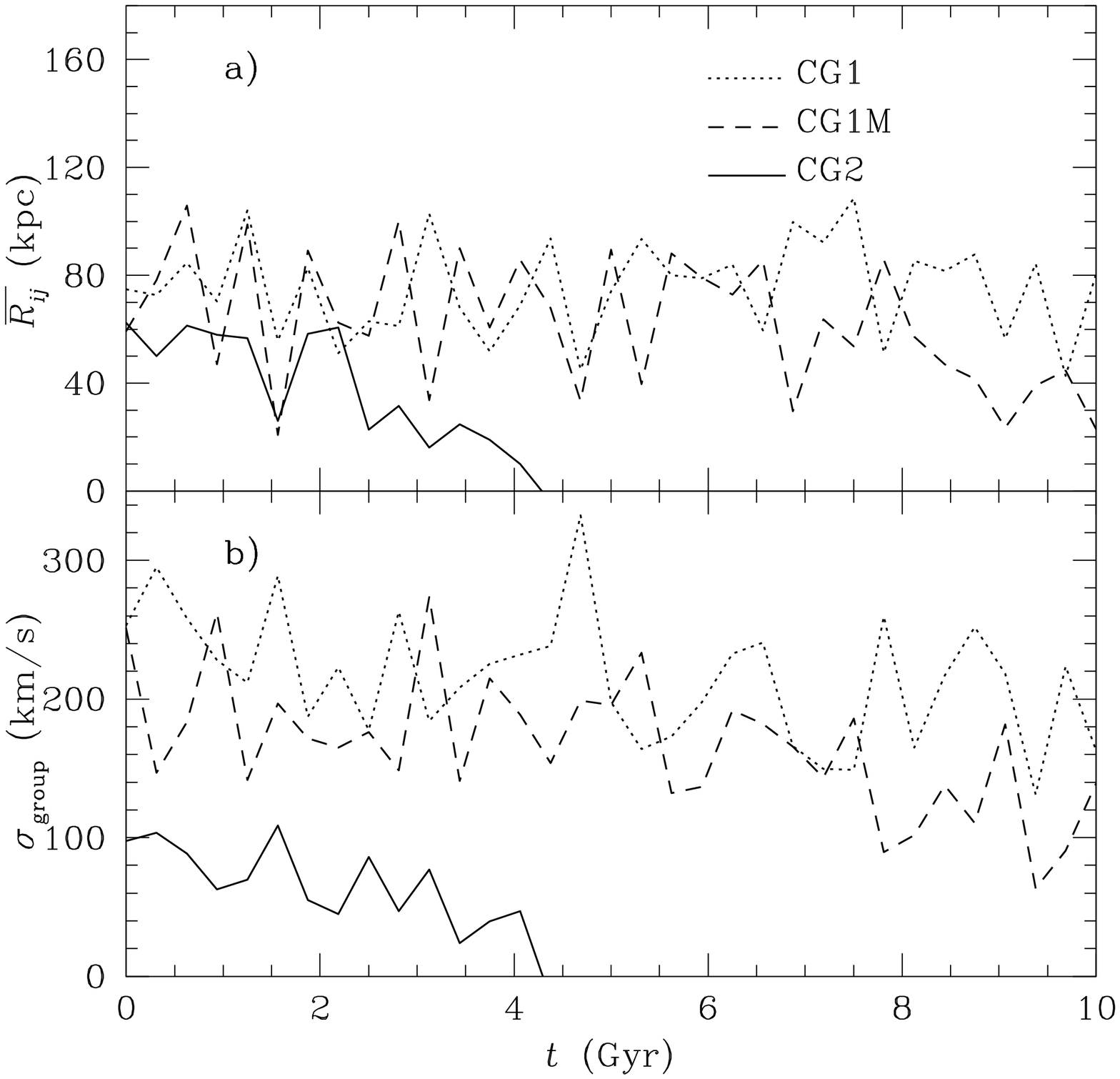}
\end{center}
\figcaption[] { (a) Evolution of the median intergalactic separation 
and (b) the group velocity dispersion
for the three group models. The merger of the remnants of
A and C galaxy models 
in the CG1M group do not affect the evolution of these group
global functions; by contrast, CG2 group disappearance is clearly seen 
in these plots.}

\end{document}